\documentclass[pra,aps,twocolumn,amssymb,nofootinbib,arajdkafj]{revtex4-1}
\usepackage{hyperref}
\usepackage{epsfig}
\begin{document}
\title{Casimir-Polder induced Rabi oscillations}
%\shorttitle{Title} %Insert here a short version of the title if it exceeds 70 characters

\author{M. Donaire}
\email[mad37ster@gmail.com, ]{donaire@lkb.upmc.fr}
\affiliation{Laboratoire Kastler Brossel, UPMC-Sorbonnes Universit\'es, CNRS, ENS-PSL Research Universities,  Coll\`{e}ge de France, 4, place Jussieu, F-75252 Paris, France}
\author{M.-P. Gorza}
\email[on leave from Laboratoire de physique des lasers, Universit\'{e} ]{
 Paris 13, Villetaneuse, France.}
\affiliation{Laboratoire Kastler Brossel, UPMC-Sorbonnes Universit\'es, CNRS, ENS-PSL Research Universities,  Coll\`{e}ge de France, 4, place Jussieu, F-75252 Paris, France}
\author{A. Maury}
\affiliation{Laboratoire Kastler Brossel, UPMC-Sorbonnes Universit\'es, CNRS, ENS-PSL Research Universities,  Coll\`{e}ge de France, 4, place Jussieu, F-75252 Paris, France}
\author{R. Gu\'{e}rout}
\affiliation{Laboratoire Kastler Brossel, UPMC-Sorbonnes Universit\'es, CNRS, ENS-PSL Research Universities,  Coll\`{e}ge de France, 4, place Jussieu, F-75252 Paris, France}
\author{A. Lambrecht}
\affiliation{Laboratoire Kastler Brossel, UPMC-Sorbonnes Universit\'es, CNRS, ENS-PSL Research Universities,  Coll\`{e}ge de France, 4, place Jussieu, F-75252 Paris, France}
%\pacs{42.50.Lc}{}
%\pacs{42.50.Ct}{}   
%\pacs{37.90.+j}{}%or 32.90.+a (interaction of atoms with photons)

\begin{abstract}
We show that the Casimir-Polder interaction may induce coherent oscillations between degenerate atomic states. We illustrate this effect by computing the Casimir-Polder induced Rabi frequency on a $^{87}$Rb atom as it interacts with a reflecting surface.
 The atom oscillates between two Zeeman sublevels of its ground state undergoing a periodic exchange of angular momentum with the vacuum photons.
\end{abstract}
\maketitle

\section{Introduction}
The interaction of a neutral atom with a material surface is a  problem profusely addressed in the literature \cite{Casimir-Polder1948,Wylie,Buhmann,Gorza,Henkel,Safari}. In most of the approaches the atom is taken in a stationary state with respect to (w.r.t) the time of observation. For distances greater than the relevant atomic transition wavelengths this interaction is referred to as \emph{retarded Casimir-Polder (CP) interaction}, while for distances much shorter than those wavelengths it is referred to as \emph{van der Waals} or \emph{non-retarded CP interaction}.

It is well known in cavity-QED that vacuum Rabi oscillations are generated  between two non-degenerate atomic states as a result of the strong coupling of the atom to a single resonant cavity mode \cite{Haroche}. In this Letter we show that Rabi oscillations  between two degenarate atomic states can also be induced by the CP interaction of the atom with a reflecting surface. In contrast to the cavity-QED vacuum Rabi oscillations, all the vacuum modes participate on the resultant \emph{CP induced Rabi oscillations}.

The net effect of the CP interaction on non-degenerate atomic states is an additive level shift  \cite{Wylie,Safari,PRADonaire2}. In particular, the resonant component of the CP interaction gives rise to an effective renormalization of the transition frequencies \cite{Safari}. On the contrary, on (quasi)degenerate atomic states the CP interaction may cause the admixture of neighboring states (eg. \cite{ScheelRibero2014}). It is in this case that, under the conditions investigated in this Letter, the atom oscillates coherently between the two (quasi)degenerate states with a well defined Rabi frequency proportional to the strength of the CP interaction. In order to prove this we study the time-evolution of the wave function of an atom initially prepared in a coherent superposition of (quasi)degenerate states. We illustrate this phenomenon with the computation of the CP induced Rabi frequency on a $^{87}$Rb atom as it interacts with a reflecting surface and oscillates between two Zeeman sublevels of its ground state. The resultant variation of atomic angular momentum is provided by the vacuum photons which mediate the interaction with the surface.

%The \revision{time-dependent} force that an excited atom experiences due to its CP interaction with a dielectric body has been considered in Ref.\cite{Safari} within a non-perturbative approach based on the density matrix formalism. The authors have found that, generically, the expression for the force cannot be derived from the spacial gradient of the generalized Lamb shift of the time-evolving atomic state. In particular, off-diagonal terms (i.e., non-additive level shifts) arise as transient states when the atom is initially prepared in a coherent superposition of eigenstates. In this approach it has been assumed that the states of the coherent superposition are non-degenerate, so that the time-evolution equations for the diagonal and off-diagonal reduced density matrix elements do not couple to each other. In this Letter we address the situation in which the initial atomic state is a coherent superposition of \emph{degenerate} or quasi-degenerate states. We use a formalism based on the calculation of the time-evolution operator to prove that the aforementioned off-diagonal terms are not transient but give rise to coherent  \emph{CP induced Rabi oscillations}.  

\section{The model}
In order to show the Rabi oscillations induced by the CP interaction of a reflecting surface on an atom, we compute the time evolution operator, $\mathbb{U}(T)$,  restricted to the two atomic states of our interest, $\{|g\rangle,|e\rangle\}$.  We assume $\omega_{eg}=\omega_{e}-\omega_{g}\geq0$.  Later, we will consider the limiting case in which $\{|g\rangle,|e\rangle\}$ forms a degenerate doublet w.r.t. the time of observation, $\omega_{eg}T\ll1$. In the following, we describe the fundamentals of the calculation.

An atom in free space with eigenstates $\{|i\rangle\}$ is described by the free Hamiltonian given by 
\begin{equation}
H_{0}^{at}=\sum_{i}\hbar\omega_{i}|i\rangle\langle i|,
\end{equation}
while the Hamiltonian of the free electromagnetic (EM) field is
\begin{eqnarray}
H_{0}^{EM}&=&\frac{1}{2}\int\textrm{d}^{3}r[\epsilon_{0}\mathbf{E}^{2}(\mathbf{r})+\mu_{0}^{-1}\mathbf{B}^{2}(\mathbf{r})]\nonumber\\
&=&\sum_{\mathbf{k},\mathbf{\epsilon}}\hbar\omega(a_{\mathbf{k},\mathbf{\epsilon}}a^{\dagger}_{\mathbf{k},\mathbf{\epsilon}}+1/2),
\end{eqnarray}
where $\mathbf{E}$ and $\mathbf{B}$ are the electric and magnetic vacuum fields respectively, $\omega=ck$ is the photon frequency, and the operators $a^{\dagger}_{\mathbf{k},\mathbf{\epsilon}}$ and $a_{\mathbf{k},\mathbf{\epsilon}}$ are the creation and annihilation operators of photons with momentum $\mathbf{k}$ and polarization $\mathbf{\epsilon}$ respectively.
In order to simplify matters the center of mass of the atom is considered fixed at a given location $\mathbf{R}$ such that, in application of the Born-Oppenheimer approximation, we can disregard the dynamics of the external atomic degrees of freedom. 

We expand the interaction potential of the atom with the EM field in a multipolar series and truncate that series at the magnetic dipole order \cite{Power},
\begin{eqnarray}
W&\simeq& W_{el}+ W_{m}, \textrm{ where}\label{W}\\
W_{el}&=&-\mathbf{d}\cdot\mathbf{E}(\mathbf{R}),\:W_{m}=-\mathbf{m}\cdot\mathbf{B}(\mathbf{R}).\nonumber
\end{eqnarray}
In these equations $\mathbf{d}$ and $\mathbf{m}$ are the atomic electric and magnetic dipole operators respectively and $\mathbf{E}(\mathbf{R})$, $\mathbf{B}(\mathbf{R})$ are the electric and magnetic vacuum field operators at the location of the atom. In the following we adopt an approach based on the fluctuation-dissipation theorem (FDT) according to which the atom is treated as a quantum system while the conducting surface is regarded as a classical object concerning its interaction with the EM field, and we restrict ourselves to zero temperature. The quantum electric and magnetic fields in the interaction potential $W$,
$\mathbf{E}$, $\mathbf{B}$, are Schr\"odinger operators and the EM vacuum state, $|\tilde{0}\rangle$, is 'dressed' by
the electric current fluctuations on the surface \cite{PRADonaire2}.  By 'dressed' we mean that the linear response of the EM field--i.e., its Green function, includes the scattering with the surface.

\begin{figure}[htb]
\includegraphics[height=8.3cm,width=7.8cm,clip]{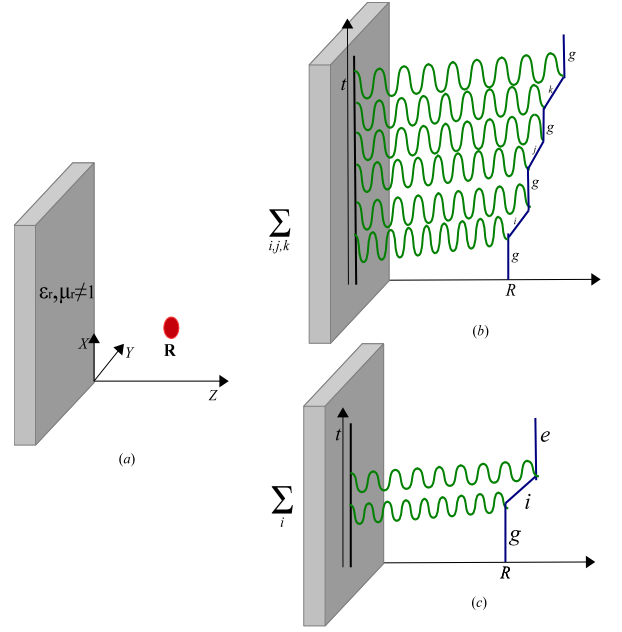}
\caption{ ($a$) Atom (in red) placed at $\mathbf{R}$ in front of a plate perpendicular to the $\hat{\mathbf{Z}}$ axis, with relative permittivity $\epsilon_{r}$ and permeability $\mu_{r}$ different to unity.  ($b$) Feynman diagram of a multiple reflection process which contributes to $_{g}\textrm{U}^{W}_{g}$ in (quasi)degenerate conditions, $\omega_{eg}T\ll1$.
This diagram is proportional to  the product of two one-particle-irreducible non-additive phase shift factors and one additive phase shift factor,
$\delta E^{\mathcal{E}}_{ge}\delta E^{\mathcal{E}}_{ee}\delta E^{\mathcal{E}}_{eg}$ --dissipative factors are omitted here for brevity. ($c$) Feynman diagram of a multiple reflection process which contributes to $_{g}\textrm{U}^{W}_{e}$ in (quasi)degenerate conditions. This diagram is proportional to the product of  three one-particle-irreducible non-additive phase shift factors, $\delta E^{\mathcal{E}}_{ge}\delta E^{\mathcal{E}}_{eg}\delta E^{\mathcal{E}}_{ge}$.  The states $|i,j,k\rangle$ are intermediate atomic states.}\label{fig3}
\end{figure}

The electric and magnetic fields can be decomposed respectively in terms of $\omega$ modes as
\begin{eqnarray}
\mathbf{E}(\mathbf{R})&=&\int_{0}^{\infty}\textrm{d}\omega[\hat{\mathbf{E}}(\mathbf{R};\omega)+\hat{\mathbf{E}}^{\dagger}(\mathbf{R};\omega)],\nonumber\\
\mathbf{B}(\mathbf{R})&=&\int_{0}^{\infty}\textrm{d}\omega[\hat{\mathbf{B}}(\mathbf{R};\omega)+\hat{\mathbf{B}}^{\dagger}(\mathbf{R};\omega)],\nonumber
\end{eqnarray}
where $\hat{\mathbf{E}}(\mathbf{R};\omega)$ $[\hat{\mathbf{B}}(\mathbf{R};\omega)]$ and $\hat{\mathbf{E}}^{\dagger}(\mathbf{R};\omega)$ $[\hat{\mathbf{B}}^{\dagger}(\mathbf{R};\omega)]$ are the creation and annihilation electric (magnetic) field operators of photons
of energy $\hbar\omega$ at the position of the atom, $\mathbf{R}$.
The quadratic vacuum fluctuations of $\hat{\mathbf{E}}(\mathbf{R};\omega)$ and $\hat{\mathbf{B}}(\mathbf{R};\omega)$  satisfy the FDT relations at zero temperature \cite{Agarwal1},
\begin{eqnarray}
\langle\tilde{0}|\hat{\mathbf{E}}(\mathbf{R};\omega)\otimes\hat{\mathbf{E}}^{\dagger}(\mathbf{R}';\omega)|\tilde{0}\rangle&=&\frac{-\hbar\omega^{2}}{\pi\epsilon_{0}c^{2}}
\Im{[\mathbb{G}(\mathbf{R},\mathbf{R}';\omega)]},\nonumber\\
\langle\tilde{0}|\hat{\mathbf{B}}(\mathbf{R};\omega)\otimes\hat{\mathbf{B}}^{\dagger}(\mathbf{R}';\omega)|\tilde{0}\rangle&=&\frac{\hbar}{\pi\epsilon_{0}c^{2}}\nonumber\\
&\times&\Im{[\nabla_{R}\wedge\mathbb{G}(\mathbf{R},\mathbf{R}';\omega)\wedge\nabla_{R'}}],\nonumber
\end{eqnarray}
where both $\mathbf{R}$ and $\mathbf{R}'$ lie on the r.h.s. of the surface where the atom is placed, $\Im$ denotes the imaginary part and $\mathbb{G}(\mathbf{R},\mathbf{R}';\omega)$ is the Green function
of the Maxwell equation for the EM field,
\begin{eqnarray}
[\frac{\omega^{2}}{c^{2}}\mathbb{\epsilon}_{r}\cdot&-&\mathbf{\nabla}\wedge(\mathbb{\mu}_{r}^{-1}\cdot\mathbf{\nabla})\wedge]\mathbb{G}(\mathbf{R},\mathbf{R}';\omega)
=\delta^{(3)}(\mathbf{R},\mathbf{R}')\mathbb{I},\nonumber\\
Z,Z'&>&0.\label{Maxwelleq}
\end{eqnarray}
In this equation $\mathbf{\epsilon}_{r}$ and $\mathbf{\mu}_{r}$ are the relative electric permittivity and magnetic permeability tensors respectively, and the atom's position lies to the right of the surface, $Z>0$ [Fig.\ref{fig3}($a$)].
The Green function can be decomposed into a free-space component and a scattering component. The contribution of the free space term to the Casimir energy
is the ordinary free-space Lamb-shift that we consider included in the bare values of the atomic transition frequencies.

\section{Generic result}

We consider $W$ as a small perturbation w.r.t. $H_{0}^{at}$ and compute the time evolution operator in the subspace $\{|g\rangle,|e\rangle\}$, $\mathbb{U}(T)$, as a power series on $W$. In the first place, the unperturbed time-evolution operator of the  atomic states $\{|i\rangle\}$, including  $|g\rangle$ and $|e\rangle$, is
\begin{equation}
\mathbb{U}^{at}_{0}(t)=\sum_{i}e^{-i\omega_{i}t}|i\rangle\otimes\langle i|.\label{Uat}
\end{equation}
On the other hand, free one-photon states evolve according to the evolution operator associated to $H_{0}^{EM}$,
\begin{equation}
\mathbb{U}^{\gamma}(t)=\sum_{\gamma_{\mathbf{k},\mathbf{\epsilon}}}e^{-i\omega t}|\gamma_{\mathbf{k},\mathbf{\epsilon}}\rangle\otimes\langle \gamma_{\mathbf{k},\mathbf{\epsilon}}|.\label{photon}
\end{equation}
If the atomic wave function at $t=0$ is $\Psi(0)=a_{g}(0)|g\rangle+a_{e}(0)|e\rangle$,  at a later time $T>0$ it reads
\begin{eqnarray}
\Psi(T)=\mathbb{U}(T)\Psi(0)&=&[ _{g}\textrm{U}_{g}(T)a_{g}(0)+\: _{g}\textrm{U}_{e}(T)a_{e}(0)]|g\rangle\nonumber\\
&+&[ _{e}\textrm{U}_{g}(T)a_{g}(0)+\: _{e}\textrm{U}_{e}(T)a_{e}(0)]|e\rangle.\nonumber
\end{eqnarray}
Straightforward application of time-dependent perturbation theory \cite{Sakurai} projected on the subspace $\{|e\rangle,|g\rangle\}\otimes|\tilde{0}\rangle$ yields the following expression for the time-evolution operator,
\begin{eqnarray}
\mathbb{U}(T)&=&[|\tilde{0}\rangle\langle\tilde{0}|\otimes(|g\rangle\langle g|+|e\rangle\langle e|)]\cdot[\mathbb{U}^{at}_{0}\otimes\mathbb{U}^{\gamma}](T)\nonumber\\
&\cdot&
\textrm{T-exp}\int_{0}^{T}\textrm{d}t[\mathbb{U}^{at}_{0}\otimes\mathbb{U}^{\gamma}]^{\dagger}(t)\cdot W\cdot[\mathbb{U}^{at}_{0}\otimes\mathbb{U}^{\gamma}](t)\nonumber\\
&\cdot&[(|g\rangle\langle g|+|e\rangle\langle e|)\otimes|\tilde{0}\rangle\langle\tilde{0}|],\label{Texp}
\end{eqnarray}
where the operators flanking $[\mathbb{U}_{0}^{at}\otimes\mathbb{U}^{\gamma}]\cdot$T-exp$\{...\}$ account for the projection onto the subspace
$\{|g\rangle,|e\rangle\}\otimes|\tilde{0}\rangle$.
At leading order the above expression contains terms proportional to the level shifts of the states $|g\rangle$ and $|e\rangle$, $\langle \tilde{g}|W|\tilde{g}\rangle$ and $\langle \tilde{e}|W|\tilde{e}\rangle$ respectively;
and  crossed terms proportional to $\langle \tilde{g}|W|\tilde{e}\rangle$ and $\langle\tilde{e}|W|\tilde{g}\rangle$, where the tilded states are states
dressed by the action of $W$. We will refer to the first kind of terms as \emph{additive} and to the latter kind as \emph{non-additive},
since only the integration of the former consists of their simple addition to the atomic energy levels. %\footnote{The non-additivity of these phases here refers to the impossibility to be attributed to shifts of internal atomic levels, while in Ref.\cite{Paulo} it refers to the impossibility to asign to them the phase shifts of individual wave packets.}. 
It can be verified that in the non-degenerate case, $\omega_{eg}T\gg1$, the contribution of the non-additive terms is of the order of $\langle\tilde{e}|W|\tilde{g}\rangle/\hbar\omega_{eg}\ll1$ and thus neglegible, so that the net effect of the CP interaction reduces here to a shift of the energy levels of the atom in the diagonal components of $\mathbb{U}$.

On the contrary, in the (quasi)degenerate case, $\omega_{eg}T\ll1$, the non-additive terms become as relevant as the additive ones. The diagrams which weight the most in the perturbative series, both in the diagonal and in the off-diagonal  components of $\mathbb{U}$, are those in which the atom transits through intermediate states with an only virtual photon before arriving repeatedly to the states $g$ and/or $e$ with no photon. %Note that in all those diagrams virtual photons do not cross each other in time, meaning that the intermediate states possess  only one virtual photon. 
Typical diagrams which contribute to diagonal and off-diagonal components of $\mathbb{U}$ are depicted in Figs.\ref{fig3}($b$) and ($c$) respectively. Their summation yields series of products of one-particle-irreducible factors, with real and imaginary (dissipative) components. Following the nomenclature explained earlier, we distinguish between additive terms, $\delta E_{gg}^{\mathcal{E}}-i\Gamma_{gg}^{\mathcal{E}}/2$, $\delta E_{ee}^{\mathcal{E}}-i\Gamma_{ee}^{\mathcal{E}}/2$;  and non-additive terms, $\delta E_{ge}^{\mathcal{E}}-i\Gamma_{ge}^{\mathcal{E}}/2$, $\delta E_{eg}^{\mathcal{E}}-i\Gamma_{eg}^{\mathcal{E}}/2$, with 
\begin{eqnarray}
\delta E_{gg}^{\mathcal{E}}&=&-\sum_{ i,\gamma_{\mathbf{k},\mathbf{\epsilon}}}
\frac{|\langle i,\gamma|W| g,\tilde{0}\rangle|^{2}}{\hbar\omega+\hbar\omega_{i\mathcal{E}}},\label{Eelggneqe}\\
\Gamma^{\mathcal{E}}_{gg}&=&\frac{2\pi}{\hbar^{2}}\sum_{i,\gamma_{\mathbf{k},\mathbf{\epsilon}}}
\Theta(\omega_{\mathcal{E}i})|\langle i,\gamma|W|g,\tilde{0}\rangle|^{2}\delta(\omega_{\mathcal{E}i}-\omega),\nonumber\\
\delta E^{\mathcal{E}}_{ge}&=&-\sum_{ i,\gamma_{\mathbf{k},\mathbf{\epsilon}}}
\frac{\langle g,\tilde{0}|W| i,\gamma\rangle\langle i,\gamma|W| e,\tilde{0}\rangle}{\hbar\omega+\hbar\omega_{i\mathcal{E}}},\label{Egee}\\
\Gamma^{\mathcal{E}}_{ge}&=&\frac{2\pi}{\hbar^{2}}\sum_{i,\gamma_{\mathbf{k},\mathbf{\epsilon}}}\Theta(\omega_{\mathcal{E}i})\langle g,\tilde{0}|W|i,\gamma\rangle
\langle i,\gamma|W|e,\tilde{0}\rangle\nonumber\\
&\times&\delta(\omega_{\mathcal{E}i}-\omega),\nonumber
\end{eqnarray}
where we have used $\mathcal{E}\equiv(E_{g}+E_{e})/2$, $\omega_{\mathcal{E}}\equiv\mathcal{E}/\hbar$. 
Analogous expressions hold for $\delta E^{\mathcal{E}}_{ee}$, $\delta E^{\mathcal{E}}_{eg}$, $\Gamma^{\mathcal{E}}_{ee}$ and $\Gamma^{\mathcal{E}}_{eg}$ with the substitution $e\leftrightarrow g$.
The single superscript $\mathcal{E}$ in the expressions for $\Gamma$ and $\delta E$  denotes the reference frequency for the transitions within the sums. Double subscripts, $gg$, $ee$, $eg$ or $ge$, denote the bra and ket states in the quantum amplitudes.

From the diagrams of Fig.\ref{fig3} we observe that each factor $(\delta E^{\mathcal{E}}_{eg}-i\hbar\Gamma^{\mathcal{E}}_{eg}/2)$ flips the state of the atom from $|e\rangle$ to $|g\rangle$, while each transposed factor produces an opposite flip. This is analogous to the action of the two Raman lasers which drive the Rabi oscillations of an atom \cite{JaynesCummings1963}. In the rotatory frame, taking the rotatory-wave-appriximation, the well-known effective Rabi Hamiltonian reads, 
\begin{equation}
\frac{\hbar}{2}\Omega|g\rangle\langle e|+\frac{\hbar}{2}\Omega^{*}|e\rangle\langle g|.
\end{equation}
The effect of the factors  $\hbar\Omega/2$ and $\hbar\Omega^{*}/2$ is analogous to that of $\delta E^{\mathcal{E}}_{ge}-i\hbar\Gamma^{\mathcal{E}}_{ge}/2$ and $\delta E^{\mathcal{E}}_{eg}-i\hbar\Gamma^{\mathcal{E}}_{eg}/2$ respectively, 
 except for the fact that in the latter case the dissipative terms break the time reversal symmetry. 
As a matter of fact, $\mathbb{U}$ can be recast in the familiar form \cite{JaynesCummings1963,PRA},  
\begin{eqnarray}
_{g}\textrm{U}_{g}(T)&=&e^{-i(\tilde{\omega}_{g}-\tilde{\Delta}/2)T}[\cos{(\Omega_{R}T/2)}-i\frac{\tilde{\Delta}}{\Omega_{R}}\sin{(\Omega_{R}T/2)}],\nonumber\\
_{g}\textrm{U}_{e}(T)&=&-ie^{-i(\tilde{\omega}_{g}-\tilde{\Delta}/2)T}\frac{|\Omega|}{\Omega_{R}}\sin{(\Omega_{R}T/2)},\nonumber\\
_{e}\textrm{U}_{e}(T)&=&e^{-i(\tilde{\omega}_{e}+\tilde{\Delta}/2)T}[\cos{(\Omega_{R}T/2)}+i\frac{\tilde{\Delta}}{\Omega_{R}}\sin{(\Omega_{R}T/2)}],\nonumber\\
_{e}\textrm{U}_{g}(T)&=&-ie^{-i(\tilde{\omega}_{e}+\tilde{\Delta}/2)T}\frac{|\Omega|}{\Omega_{R}}\sin{(\Omega_{R}T/2)},\label{UR}
\end{eqnarray}
with the  parameters so defined \footnote{Note here
that $\Omega^{*}$ so defined is not the complex conjugate of $\Omega$ because of the damping factors.},
\begin{eqnarray}
\tilde{\omega}_{g}&\equiv&\omega_{g}+\delta E^{\mathcal{E}}_{gg}/\hbar-i\Gamma^{\mathcal{E}}_{gg}/2,\nonumber\\
\tilde{\omega}_{e}&\equiv&\omega_{e}+\delta E^{\mathcal{E}}_{ee}/\hbar-i\Gamma^{\mathcal{E}}_{ee}/2,\qquad
\tilde{\Delta}\equiv\tilde{\omega}_{e}-\tilde{\omega}_{g},\nonumber\\
\Omega&\equiv&2\delta E^{\mathcal{E}}_{ge}/\hbar-i\Gamma^{\mathcal{E}}_{ge},
\quad\Omega^{*}\equiv 2\delta E^{\mathcal{E}}_{eg}\hbar-i\Gamma^{\mathcal{E}}_{eg},\nonumber\\
|\Omega|^{2}&\equiv&\Omega\Omega^{*},\quad\Omega_{R}\equiv\sqrt{|\Omega|^{2}+\tilde{\Delta}^2}.\nonumber
\end{eqnarray}
This means that the Casimir-Polder interaction may indeed induce  Rabi oscillations  between two quiasi-degenerate states with a \emph{CP induced Rabi frequency}
 $\Omega_{R}=\sqrt{|\Omega|^{2}+\tilde{\Delta}^{2}}$.\\

 \section{Application: CP induced Rabi frequency on a $^{87}$Rb atom}
 We propose here a simple setup which
illustrates the phenomenon of interest. We study the oscillations of a $^{87}$Rb atom  between
 two degenerate Zeeman sublevels of its ground state,
 $|g\rangle=|5^{2}S_{1/2},F=1,m_{F}=-1\rangle$ and $|e\rangle=|5^{2}S_{1/2},F=1,m_{F}=+1\rangle$, at zero temperature in the vicinity of a perfectly reflecting metallic surface.  The CP induced level shifts of  Zeeman states on a trapped electron have been studied recently in Ref.\cite{Carla}. For reasons that will be clear later, we must take the axis of quantization parallel to the surface. If the surface is perpendicular to the axis $\hat{\mathbf{Z}}$, we choose the quantization axis along the $\hat{\mathbf{X}}$ direction.

The non-additive phase shift term is largely dominated by the electric dipole interaction. Making use of the FDT we find, as a function of the dyadic Green's function \cite{Wylie,Buhmann,PRADonaire2},
\begin{eqnarray}
\delta E^{\mathcal{E}}_{ge}&=&\frac{-1}{\pi\epsilon_{0}c^{2}}\sum_{i}\int_{0}^{\infty}\textrm{d}u  \frac{u^{2}\omega_{i\mathcal{E}}}{u^{2}+\omega_{i\mathcal{E}}^{2}}\nonumber\\
&\times&\textrm{Tr}\{\langle g|\mathbf{d}|i\rangle\cdot\mathbb{G}(\mathbf{R},\mathbf{R};iu)\cdot\langle i|\mathbf{d}|e\rangle\}\label{dEge}\\
&+&\frac{1}{\epsilon_{0}c^{2}}\sum_{i}
\Theta(\omega_{\mathcal{E}i})\omega_{\mathcal{E}i}^{2}\nonumber\\&\times&\textrm{Tr}\{\langle g|\mathbf{d}|i\rangle
\cdot\Re[\mathbb{G}(\mathbf{R},\mathbf{R};\omega_{\mathcal{E}i})]\cdot\langle i|\mathbf{d}|e\rangle\},\label{dEge2}
\end{eqnarray}
where the first term corresponds to the off-resonant component ($iu$ is the complex frequency) and the second one is the resonant component. In the basis of hyperfine states labeled with the quantum numbers $F$ and $m_{F}$, it is convenient to express the dipole moment operator in the spherical basis,
\begin{equation}
d_{y}=(d_{-}-d_{+})/\sqrt{2},\: d_{z}=i(d_{-}+d_{+})/\sqrt{2},\:d_{x}=d_{0},
\end{equation}
where all the dipole transition matrix elements are real numbers. In this basis we find,
\begin{eqnarray}\label{Gs}
&\textrm{Tr}\{\langle g|\mathbf{d}|i\rangle\cdot\mathbb{G}\cdot\langle i|\mathbf{d}|e\rangle\}=
G_{xx}\langle g|d_{0}|i\rangle\langle e|d_{0}|i\rangle\\&+
(G_{zz}-G_{yy})(\langle g|d_{+}|i\rangle\langle e|d_{-}|i\rangle+\langle g|d_{-}|i\rangle\langle e|d_{+}|i\rangle)/2\nonumber\\&+
(G_{zz}+G_{yy})(\langle g|d_{+}|i\rangle\langle e|d_{+}|i\rangle+\langle g|d_{-}|i\rangle\langle e|d_{-}|i\rangle)/2,\nonumber
\end{eqnarray}
where the frequency and position dependence of the dyadic Green function components have been omitted. The off-diagonal terms of the Green function are identically zero. We find for a perfectly conducting reflector \cite{Wylie,Gorza,Buhmann},
 \begin{eqnarray}\label{Geeny}
G_{xx}(Z;\omega)=G_{yy}(Z;\omega)&=&\frac{-e^{2ikZ}}{32\pi k^{2}Z^{3}}(1-2ikZ-4k^{2}Z^{2}),\nonumber\\
G_{zz}(Z;\omega)&=&\frac{e^{2ikZ}}{16\pi k^{2}Z^{3}}(-1+2ikZ).
\end{eqnarray}
Straightforward application of selection rules \cite{SeckRb87}  in Eq.(\ref{Gs}) implies that only 
the term of Eq.(\ref{Gs}) proportional to $G_{zz}-G_{yy}$ may survive, which explains why the axis of quantization must be taken parallel to the surface. 

\begin{figure}[htb]
\includegraphics[height=12.0cm,width=8.3cm,clip]{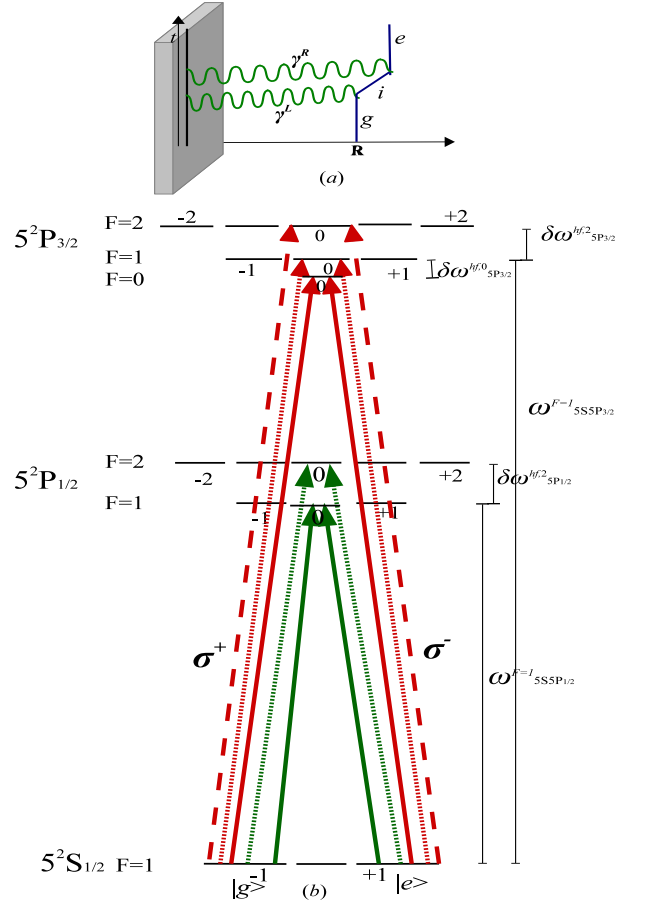}
\caption{($a$) Feynman diagram of the non-additive  term $\delta E^{\mathcal{E}}_{ge}$. Initially, virtual photons of  left handed circular polarization, $\gamma^{L}$, are created at the time the atom is in state $|g\rangle$. Later, $\gamma^{L}$ photons get reflected off the surface turning into right handed circularly polarized, $\gamma^{R}$. Finally, $\gamma^{R}$ photons are annihilated at the time the atom gets to state $|e\rangle$. ($b$) Schematic representation of the virtual transitions which drives a $^{87}$Rb atom through coherent oscillations between two Zeeman sublevels of its 
ground state, $|g\rangle=|5^{2}S_{1/2},F=1,m_{F}=-1\rangle$ and $|e\rangle=|5^{2}S_{1/2},F=1,m_{F}=+1\rangle$.  Virtual photons of opposite circular polarization mediate the 
transition of $|g\rangle$ and $|e\rangle$,  $\sigma^{+}$ and $\sigma^{-}$ transitions respectively, to a series of common intermediate states --states $|i\rangle$ in ($a$)-- which are hyperfine Zeeman sublevels of $5^{2}P_{1/2}$ (in green) and $5^{2}P_{3/2}$ (in red) with $m_{F}=0$.  The hyperfine levels with $F=1$ are considered as the reference levels. The transition frequencies are \cite{SeckRb87}, $\omega_{5S5P_{1/2}}^{F=1}=2\pi377.11$THz, $\omega_{5S5P_{3/2}}^{F=1}=2\pi384.23$THz,  and  the hyperfine intervals, $\delta\omega^{hf,2}_{5P_{1/2}}=2\pi0.817$GHz, $\delta\omega^{hf,2}_{5P_{3/2}}=2\pi0.157$GHz, $\delta\omega^{hf,0}_{5P_{3/2}}=-2\pi0.072$GHz.}\label{fig4}
\end{figure}

In the calculation of $\delta E^{\mathcal{E}}_{ge}$ it suffices to consider the D1 and D2 line transitions, $5^{2}S_{1/2}\rightarrow5^{2}P_{1/2}$ and $5^{2}S_{1/2}\rightarrow5^{2}P_{3/2}$ respectively.  In Fig.\ref{fig4}($b$) it appears depicted all possible virtual transitions from the states $|g\rangle$ and $|e\rangle$ which contribute to $\delta E^{\mathcal{E}}_{ge}$ and $\delta E^{\mathcal{E}}_{eg}$. In all of them the intermediate states are hyperfine Zeeman sublevels of $5^{2}P_{1/2}$ and $5^{2}P_{3/2}$ with $m_{F}=0$, and the virtual photons created and annihilated during the transitions possess opposite circular polarization, $\gamma^{L,R}$ and $\gamma^{R,L}$ respectively [see Fig.\ref{fig4}($a$)]. The latter implies a net variation of $2\hbar$ in the atomic angular momentum. Important is the fact that a non-zero value of $\delta E^{\mathcal{E}}_{ge}$ needs the inclusion of retardation in the electric Casimir interaction, even at submicron distances. The reason being that the following relation holds, for $j=1/2,3/2$,
\begin{equation}
 \sum_{F=|j-3/2|}^{j+3/2}\langle e|d_{-}|5^{2}P_{j},F,m_{F}=0\rangle\langle g|d_{+}|5^{2}P_{j},F,m_{F}=0\rangle=0.\nonumber
\end{equation}
Therefore, non-zero hyperfine intervals in the states $|5^{2}P_{j}\rangle$, $\delta\omega^{hf,F}_{5P_{j}}\equiv\omega^{F}_{5S5P_{j}}-\omega^{F=1}_{5S5P_{j}}$, are necessary for $\delta E^{\mathcal{E}}_{ge}$ not to vanish.  Inserting the formula of Eq.(\ref{Gs}) into Eq.(\ref{dEge}) we get,
\begin{eqnarray}
\delta E^{\mathcal{E}}_{eg}&\simeq&\frac{-1}{32c\pi^{2}\epsilon_{0}Z^{2}}\int_{0}^{\infty}\textrm{d}u\frac{f(u)}{u^{2}+\kappa_{jF}^{2}}\sum_{j=1/2,F=|j-3/2|}^{3/2,j+3/2}\omega^{F}_{5S5P_{j}}\nonumber\\
&\times&\langle e|d_{-}|5^{2}P_{j},F,m_{F}=0\rangle
\langle g|d_{+}|5^{2}P_{j},F,m_{F}=0\rangle,\nonumber
\end{eqnarray}
 where  $f(u)=e^{-2u}(1+2u-4u^{2})$ and $\kappa_{jF}=\omega^{F}_{5S5P_{j}}Z/c$, with  $\omega^{F}_{5S5P_{j}}$ being the transition frequency from the hyperfine level $F$ of  $5^{2}P_{j}$ to the states $|g\rangle$, $|e\rangle$. 
 At leading order in $\delta\omega^{hf,F}_{5P_{j}}$ we find
 \begin{eqnarray}
\delta E^{\mathcal{E}}_{eg}&\simeq&\sum_{j=1/2,F=|j-3/2|}^{3/2,j+3/2}\frac{-|\langle 5^{2}S_{1/2}||\mathbf{d}||5^{2}P_{j}\rangle|^{2}(\omega^{F=1}_{5S5P_{j}})^{2}}{2^{(j-1/2)(F-1)}384\pi^{2}c^{3}\epsilon_{0}}\nonumber\\
&\times&\delta\omega^{hf,F}_{5P_{j}}\int_{0}^{\infty}\textrm{d}u\Bigl[\frac{\kappa_{j1}^{-2}f(u)}{u^{2}+\kappa_{j1}^{2}}
-\frac{2f(u)}{(u^{2}+\kappa_{j1}^{2})^{2}}\Bigr],\label{derrier}
\end{eqnarray}
where $\langle 5^{2}S_{1/2}||\mathbf{d}||5^{2}P_{1/2,3/2}\rangle$ are the reduced dipole moment matrix elements of the D$_{1}$ and D$_{2}$ transition lines respectively.

\begin{figure}[htb]
\includegraphics[height=11.cm,width=8.6cm,clip]{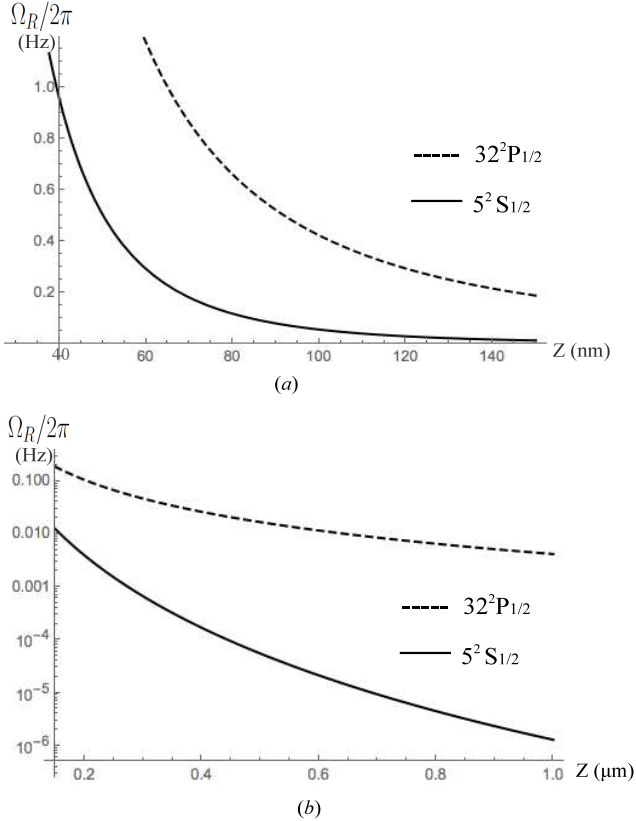}
\caption{Graphical representation of the values of  $\Omega_{R}/2\pi=2\delta E^{\mathcal{E}}_{eg}/h$ given by Eqs.(\ref{dEge},\ref{dEge2},\ref{derrier}) in the ranges $(a)$ $40$nm$\leq Z\leq150$nm and $(b)$ $0.15\mu$m$\leq Z\leq1.0\mu$m. The continuous lines correspond to the case where $|g\rangle$ and $|e\rangle$ belong to the ground state $5^{2}S_{1/2}$. The dashed lines  correspond to the case where $|g\rangle$ and $|e\rangle$ belong to the Rydberg state $32^{2}P_{1/2}$.}
\label{fig5}
\end{figure}

Tabulated data of all the unknowns in Eq.(\ref{derrier}) can be found in Ref.\cite{SeckRb87}.  Fig.\ref{fig5} contains the plot of $\Omega_{R}/2\pi=2\delta E_{eg}^{\mathcal{E}}/h$ as a function of the distance to the surface. In particular, for $|g\rangle$ and $|e\rangle$ in $5^{2}S_{1/2}$, the  CP induced Rabi frequency is $2\pi$Hz at 40nm and $2\pi$mHz at 270nm approximately.  It is worth stressing that the symmetry of the setup yields the same additive shifts for the states $|g\rangle$ and $|e\rangle$, keeping this way their degeneracy. In particular, the electric CP interaction provides a level shift  between 0.2MHz and 70 MHz within the interval 40nm$\leq Z\leq270$nm, which guarantees the perturbative nature of the calculation; and the magnetic interaction of the net magnetic moments of the states $|g\rangle$ and $|e\rangle$, $\pm(\mu_{B}/2)\hat{\mathbf{X}}$ respectively, provides them with a common shift in the range 0.01Hz-4.3Hz which raises them over the state $|5^{2}S_{1/2},F=1,m_{F}=0\rangle$. This might be a potential problem for the observation of Rabi oscillations  since both $|g\rangle$ and $|e\rangle$ can
decay into the state $|5^{2}S_{1/2},F=1,m_{F}=0\rangle$ through an M1 transition with the same rate, $\Gamma^{\mathcal{E}}_{gg}=\Gamma^{\mathcal{E}}_{ee}$. This rate depends on the actual conductivity of the reflecting surface through
its skin depth, $\xi$. For the reasonable assumption that $\xi\gg300$nm %\footnote{As an example, $\xi\simeq75\mu$m for gold, and $\xi\simeq5\mu$m for nickel.}, 
we find \cite{Henkel,PRA}, $\Gamma^{\mathcal{E}}_{gg}\simeq 20\pi$Hz(40nm$/\xi)^{2}$ at $Z=40$nm and  $\Gamma^{\mathcal{E}}_{gg}\simeq 65\pi$mHz(270nm$/\xi)^{2}$ at $Z=270$nm. In order to complete a Rabi cycle, $\Gamma^{\mathcal{E}}_{gg}<\Omega_{R}$ is necessary, which  is guaranteed by the initial assumption $\xi\gg300$nm within the interval 40nm$\leq Z\leq270$nm. In addition, we note that similar values are obtained for the non-additive dissipative terms, $\Gamma^{\mathcal{E}}_{ge,eg}$, which means that their contribution to $\Omega_{R}$  can be discarded in good approximation. 

Finally, we observe that through a Rabi semi-cycle, $\pi/\Omega_{R}$, the  total angular momentum  of the atom, $\mathbf{L}_{F}$, varies an amount $2\hbar$ along the quantization axis $\hat{\mathbf{X}}$. For an atom in the state $|g\rangle$ at $t=0$, it holds $\mathbf{L}_{F}(t)=-\hbar\cos{\Omega_{R}t}\hat{\mathbf{X}}$. More specifically, during a Rabi semi-cycle the nuclear spin varies from $-5\hbar/4$ to $+5\hbar/4$ while the electronic spin varies in the oppossite sense, from $+\hbar/4$ to  $-\hbar/4$ along $\hat{\mathbf{X}}$, such that $\Delta m_{F}=2$. This variation of angular momentum is provided by the vacuum photons which mediate the interaction with the reflecting surface as explained in Fig.\ref{fig4}($a$).  At a theoretical level, conservation of total angular momentum implies that, in order to compensate for this variation, the angular momentum of the reflecting surface must vary an equal amount in opposite sense. It explains why the axis of quantization must be taken parallel to the surface as the surface is seen as an infinite plane from the atom. The rotational invariance of an infinite plane around its perpendicular axis -- the $\hat{\mathbf{Z}}$ axis in our case -- would prevent otherwise the exchange of angular momentum between the surface and the atom in this direction. An explicit calculation of the angular momentum of the surface would require a microscopical model for the interaction between the surface and the EM field. 

We finish this section with a remark on Rydberg states.  
At first glance it could be thought that greater values of $\Omega_{R}$ would be obtained in the oscillation between Zeeman sublevels of Rydberg states with large principal number $n$. A study of the scaling behaviour as a function of $n$ of the dipole moment matrix elements and resonant frequencies which enter the expression of $\delta E_{eg}^{\mathcal{E}}$ \cite{PRA} reveals that much greater values of $\Omega_{R}$ can only be obtained at distances for which the CP interaction enters the retarded regime w.r.t. the D$_{1}$ and D$_{2}$ transtions of the ground state, while it remains in the non-retarded regime w.r.t. the transition between neighboring Rydberg states. This is illustrated in Fig.\ref{fig5}, where the Rydberg state is chosen to be $32^{2}P_{1/2}$. The calculation of $\Omega_{R}$ is analogous to the one presented above for the Zeeman sublevels of the ground state \cite{PRA}. At distances greater than $\sim150$nm, the values of $\Omega_{R}$ are much larger for oscillations between  Zeeman sublevels of the Rydberg state. However, those values are still too small to be observed due to the short lifetimes of the excited states \cite{JPB2010}.

\acknowledgments
We thank Franck Pereira dos Santos and Peter Wolf for stimulating discussions. 
Financial support from ANR-10-IDEX-0001-02-PSL and ANR-13-BS04--0003-02 is gratefully acknowledged.

\end{document}